\documentclass[preprint]{elsarticle}
\usepackage{amsmath,amsfonts,amssymb}
\usepackage{latexsym}
\usepackage{dcolumn}
\usepackage{graphicx,epsfig}
\usepackage{amsthm}
\usepackage{hyperref}

\begin{document}
\title{\bf Do the Modified Uncertainty Principle and Polymer Quantization predict same physics?}
\author{Barun Majumder}
\ead{barunbasanta@iiserkol.ac.in}
\address{Indian Institute of Science Education and Research (Kolkata) \\ Mohanpur, Nadia, West Bengal, Pin 741252\\ India}
\author{Sourav Sen}
\ead{souravsen@iiserkol.ac.in}
\address{Indian Institute of Science Education and Research (Kolkata) \\ Mohanpur, Nadia, West Bengal, Pin 741252\\ India}
\begin{frontmatter}
\begin{abstract} 
In this Letter we study the effects of the Modified Uncertainty Principle as proposed in Ali et al. (2009) \cite{b1} in simple quantum mechanical systems and study its thermodynamic properties. We have assumed that the quantum particles follow Maxwell-Boltzmann statistics with no spin. We compare our results with the results found in the GUP and polymer quantum mechanical frameworks. Interestingly we find that the corrected thermodynamic entities are exactly same compared to the polymer results but the length scale considered has a theoretically different origin. Hence we express the need of further study for an investigation whether these two approaches are conceptually connected in the fundamental level.
\end{abstract}
\end{frontmatter}

\section*{Introduction}
The idea that the uncertainty principle could be affected by gravity was first given by Mead \cite{c1}. Later modified commutation relations between position and momenta commonly
known as Generalized Uncertainty Principle ( or GUP ) were given by candidate theories of quantum gravity ( String Theory, Doubly Special
Relativity ( or DSR ) Theory and Black Hole Physics ) with the prediction of a minimum measurable length \cite{b2,b7}. Similar kind of
commutation relation can also be found in the context of Polymer Quantization in terms of Polymer Mass Scale \cite{c}.
\par
The authors in \cite{b1} proposed a MUP\footnote{from now we denote this as MUP to distinguish it from the Generalized Uncertainty Principle (GUP) introduced in \cite{kempf}} which is consistent with DSR theory, String theory and Black Hole Physics and which says
\begin{equation}
\left[x_i,x_j\right] = \left[p_i,p_j\right] = 0 ,
\end{equation}
\begin{equation}
\label{g2}
[x_i, p_j] = i \hbar \left[  \delta_{ij} -  \alpha  \left( p \delta_{ij} +
\frac{p_i p_j}{p} \right) + \alpha^2  \left( p^2 \delta_{ij}  + 3 p_{i} p_{j} \right)  \right],
\end{equation}
\begin{align}
\label{g3}
 \Delta x \Delta p &\geq \frac{\hbar}{2} \left[ 1 - 2 \alpha <p> + 4 \alpha^2 <p^2> \right]  \nonumber \\
& \geq \frac{\hbar}{2} \left[ 1  +  \left(\frac{\alpha}{\sqrt{\langle p^2 \rangle}} + 4 \alpha^2  \right)  \Delta p^2  +  4 \alpha^2 \langle p \rangle^2 -  2 \alpha \sqrt{\langle p^2 \rangle} \right],
\end{align}
where $ \alpha=\frac{l_0 l_{pl}}{\hbar} $. Here $ l_{pl} $ is the Plank length ($ \approx 10^{-35} m $). It is normally assumed that the dimensionless
parameter $l_0$ is of the order unity. If this is the case then the $\alpha$ dependent terms are only important at or near the Plank
regime. But here we expect the existence of a new intermediate physical length scale of the order of $\alpha \hbar = l_0 l_{pl}$. We also note
that this unobserved length scale cannot exceed the electroweak length scale \cite{b1} which implies $l_0 \leq 10^{17}$. These equations are
approximately covariant under DSR transformations but not Lorentz covariant \cite{b7}. These equations also imply
\begin{equation}
\Delta x \geq \left(\Delta x \right)_{min} \approx l_0\,l_{pl}
\end{equation}
and
\begin{equation}
\Delta p \leq \left(\Delta p \right)_{max} \approx \frac{M_{pl}c}{l_0}
\end{equation}
where $ M_{pl} $ is the Plank mass and $c$ is the velocity of light in vacuum. It can be shown that equation (\ref{g2}) is satisfied by the
following definitions $x_i=x_{oi}$ and $p_i=p_{oi} (1 - \alpha \,p_o + 2\,\alpha^2\,p_o^2)$, where $x_{oi}$, $p_{oj}$ satisfies $[x_{oi}, p_{oj}]= i \hbar \delta_{ij}$. Here we can interpret $p_{oi}$ as the momentum at low energies having the standard representation in position space ($ p_{oi}=-i\hbar \frac{\partial}{\partial x_{oi}}$) with $p_o^2=\sum_{i=1}^3 p_{oi}p_{oi}$ and $ p_i $ as the momentum at high energies. We can also show that the $ p^2 $ term in the kinetic part of any Hamiltonian can be written as \cite{b1}
\begin{equation}
\label{corr}
 p^2 \Longrightarrow \ p_o^2 - 2\ \alpha \ p_o^3 + {\cal O}(\alpha^2) + \ldots \, \, \, .
\end{equation}
Here we assume that terms $ {\cal O}(\alpha^2)$ are much smaller in magnitude in comparison to terms $ {\cal O}(\alpha)$ as $\alpha=l_0\,l_{pl}$. The effect of this proposed MUP is well
studied recently for some well-known physical systems in \cite{b1,z16,bound,areahole,z17,cite5}.\par
In this Letter we study two simple quantum systems (particle in a box and harmonic oscillator) in the MUP framework and calculate the thermodynamic entities
for each case. Later we compare our results with those as predicted by GUP \cite{p58} and Polymer Quantum Mechanics \cite{poly} and interestingly we found that the physical interpretation remain same provided that the polymer length scale is theoretically very different in its origin.

\section*{Particle in a box }
As we have mentioned earlier that we are going to study the MUP corrected quantum particle in a box problem (considering the quantum particle have no spin), so here we again mention the $p^{2}$ term in the Hamiltonian can be replaced by $p_o^{2}-2{\alpha}p_o^{3}+\mathcal{O}\left(\alpha^{2}\right)$ where we have already designed $p_o$ earlier and for our purpose we can use $p_{oi}{\equiv}-i{\hbar}\frac{\partial}{{\partial}x_i}$. 
\par
We can easily write the MUP corrected Scr\"{o}dinger Equation for the particle in a box (in one dimension) as
\begin{equation}
\label{e1}
{\gamma}\frac{\partial^{3}\Psi}{{\partial}x^{3}}+\frac{\partial^{2}\Psi}{{\partial}x^{2}}+k^{2}\Psi=0~~,
\end{equation}
where $\gamma=2i{\alpha}\hbar$ and $k^{2}=\frac{2mE}{\hbar^{2}}$. It is clear from equation (\ref{e1}) that if $\alpha=\gamma=0$ we get back the usual known equation with respect to some boundary condition. Here we are going to solve equation (\ref{e1}) in some perturbative sense. If $\gamma=0$ we get $\Psi\sim{\sin}k_o{x}$. The subscript {\textquoteleft{\scriptsize o}\textquoteright} refers to the normal situation where we have not considered the MUP effect and $k_o=\sqrt{\frac{2mE_o}{\hbar^{2}}}$.
\par
Now if we use the approximation $\Psi\sim{\sin}k_o{x}$ we can re-write equation (\ref{e1}) as
\begin{equation}
\label{e2}
\frac{\partial^{2}\Psi}{{\partial}x^{2}}-{\gamma}{k_o^{2}}\frac{\partial\Psi}{{\partial}x}+k^{2}\Psi=0~~,
\end{equation}
where $\lim_{\gamma \to 0} k=k_o$. We can now write the solution for $\Psi$ as
\begin{equation}
\label{e3}
\Psi{\sim}e^{i\alpha{\hbar}k_o^{2}x}{\sin}\left(\sqrt{\alpha^{2}\hbar^{2}k_o^{4}+k^{2}}x\right)~~,
\end{equation}
where we have already exploited are boundary condition $\Psi(x=0)=0$.
\par
With another condition $\Psi(x=L)=0$, where $L$ is the length of the box, we get the quantization relation
\begin{equation}
\label{e4}
\sqrt{\alpha^{2}\hbar^{2}k_o^{4}+k^{2}}~L = n\pi~~.
\end{equation}
After some straightforward rearrangement we can finally write
\begin{equation}
\label{e5}
E_n=E_{on}-2~\alpha^2 ~m ~E_{on}^2~~,
\end{equation}
where $E_{on}=\frac{n^{2}\pi^{2}\hbar^{2}}{2mL^{2}}$ is the energy eigenvalue if $\alpha=0$ (standard case).
\par
This particular problem was studied earlier \cite{p58} where the generalized commutation relation was considered to be the one well discussed in \cite{kempf}. There also we can find a correction is proportional to the square of the minimum length ($l_o$) with respect to the generalized commutation relation used. We can see
\begin{equation}
\label{e6}
E_{nGUP}=\frac{n^{2}\pi^{2}\hbar^{2}}{2mL^{2}}+l_o^{2}\frac{n^{4}\pi^{4}\hbar^{2}}{3mL^{4}}~~.
\end{equation}
\par
So now if we compare equation (\ref{e6}) with (\ref{e5}) we can see that the correction is proportional to the square of the minimum length and its coefficient is also same but there is a difference in sign in the prefactor. So the intention of the correction term is not the same in both the cases.
\par
Now this particular problem is also studied in the realm of Polymer Quantization \cite{poly} and the approximate spectrum is found out to be
\begin{equation}
\label{e7}
E_{nPoly}=\frac{n^{2}\pi^{2}\hbar^{2}}{2mL^{2}}-\lambda^{2}\frac{n^{4}\pi^{4}\hbar^{2}}{24{m}L^{4}}~~,
\end{equation}
where $\lambda=\frac{\mu_o}{L}$ and $\mu_o$ is considered as a constant but related to the polymer length scale. Now if we compare this result with equation (\ref{e5}) we see that these two corrections are similar (if we neglect the numerical factor of 24). Authors in \cite{poly} have argued that this coincidence is not surprising since polymer systems have similar modifications to GUP in their corresponding uncertainty relation \cite{p39}. But the MUP is having a term linear in Planck length in the commutation relation and still we can find corrections that are nearly or exactly similar to those as predicted by GUP \cite{p58} or Polymer Quantization \cite{poly}. The first order correction term is quadratic in length scale for all the cases but the energies are reduced both in the MUP and Polymer framework.
\par
Now we are going to apply the MUP corrected energy spectrum to calculate the canonical partition function and other thermodynamic quantities for the ideal gas assuming the quantum particles follow Maxwell-Boltzmann statistics with no spin. So we first calculate the MUP corrected partition function and it can be expressed as
\begin{equation}
\label{e8}
Z(\beta)\approx\sum_{n=1}^\infty{\exp}\left(\frac{-c{\beta}n^{2}}{L^{2}}\right)+\frac{d{\beta}{\alpha^{2}}}{L^{4}}\sum_{n=1}^\infty{n^{4}}{\exp}\left(\frac{-c{\beta}n^{2}}{L^{2}}\right)+\mathcal{O}\left(\alpha^{4}\right)~~,
\end{equation}
where $c=\frac{\pi^{2}\hbar^{2}}{2m}$, $d=\frac{\pi^{4}\hbar^{4}}{2m}$ and $\beta=\frac{1}{{K_B}T}$. From now on we are going to work in the unit where $\hbar=l_{pl}=1$.
If we apply the Poisson Summation formula \cite{wolfram} we can re-write the partition function as
\begin{equation}
\label{e9}
Z(\beta){\approx}\left(\frac{mL^{2}}{2{\pi}\beta}\right)^{\frac{1}{2}}+\alpha^{2}\left(\frac{3m}{2\beta}\right)\left(\frac{mL^{2}}{2{\pi}\beta}\right)^{\frac{1}{2}}+....\approx\left(\frac{mL^{2}}{2{\pi}\beta}\right)^{\frac{1}{2}}\left[1+\alpha^{2}\left(\frac{3m}{2\beta}\right)+....\right]~~.
\end{equation}
The first term is exactly the partition function for one-dimensional ideal gas. Now as we are using the MUP in a perturbative sense so for our approximation to be valid we require $\alpha<<\sqrt{\frac{2\beta}{3m}}$ as $\alpha=\frac{{l_o}{l_{pl}}}{\hbar}$ and $l_{pl}=\hbar=1$. For the case of indistinguishable particles we use the relation $F=-\frac{1}{\beta}\ln\left(\frac{Z^{\cal N}}{{\cal N}!}\right)$ for the Helmholtz free energy and it comes out to be
\begin{equation}
\label{e10}
F\approx\frac{-{\cal N}}{\beta}\left[1+\ln\left(\frac{L}{\cal N}\right)+\frac{1}{2}\ln\left(\frac{m}{2\pi\beta}\right)+\ln\left(1+\frac{3\alpha^{2}m}{2\beta}\right)+.....\right]~~.
\end{equation}
The chemical potential ($\mu = \frac{\partial F}{\partial {\cal N}}$) and the entropy ($S = k_B \beta^2 \frac{\partial F}{\partial \beta}$) can also be calculated
\begin{equation}
\label{e11}
\mu=\frac{-1}{\beta}\left[\ln\left(\frac{L}{\cal N}\right)+\frac{1}{2}\ln\left(\frac{m}{2\pi\beta}\right)+\ln\left(1+\frac{3\alpha^{2}m}{2\beta}\right)+.....\right]~~\&
\end{equation}
\begin{equation}
\label{e12}
S={\cal N}{K_B}\left[\frac{3}{2}+\ln\left(\frac{L}{\cal N}\right)+\frac{1}{2}\ln\left(\frac{m}{2\pi\beta}\right)+\frac{3\alpha^{2}m}{2\beta}+\ln\left(1+\frac{3\alpha^{2}m}{2\beta}\right)+.....\right]~~.
\end{equation}
Equation (\ref{e12}) is the MUP modified {\sl Sackur-Tetrode equation}. The internal energy ($U = -\frac{{\cal N}}{Z} \frac{\partial Z}{\partial \beta}$) and the heat capacity ($C_V = -k_B \beta^2 \frac{\partial U}{\partial \beta}$) are respectively
\begin{equation}
\label{e13}
U=\frac{\cal N}{2\beta}\left(1+\frac{3\alpha^{2}m}{\beta}+...\right)~~\&
\end{equation}
\begin{equation}
\label{e14}
C_V=\frac{{\cal N}K_B}{2}\left(1+\frac{6\alpha^{2}m}{\beta}+.....\right)~~.
\end{equation}
\par
It is very important to note that these thermodynamic quantities (\ref{e10}-\ref{e14}) were also extensively studied from the perspective of Polymer Quantization \cite{poly}. The correction terms that we have found are exactly similar (upto numerical prefactors) to those found in \cite{poly} but the length scale is different. But if we compare our results with those calculated from GUP we can see that in our case, the corrections tend to increase the thermodynamic quantities. The GUP motivated corrections decrease the quantities.We have noticed that our approximation is valid when $\alpha<<\sqrt{\frac{2\beta}{3m}}$. In equation (\ref{e10}) we see that $F$ diverges as $\beta{\rightarrow}0$ or $T{\rightarrow}\infty$. We should avoid $\beta{\rightarrow}0$ limit as this is beyond our approximation.

\section*{Harmonic Oscillator}
In this section of the Letter, we will study the effect of the MUP in the context of Harmonic Oscillator. Our analysis will be perturbative as in the first approximation we will neglect terms ${\cal O}(\alpha^2)$. For our purpose, we now take a different route and consider the modified Heisenberg algebra \cite{z17} where $\bf{x}$ and $\bf{p}$ obeys the relation ($a >0$)
\begin{equation}
\label{ee12}
[\mathbf{x}~,~\mathbf{p}] = i~(1-a~\mathbf{p})~~.
\end{equation}
We have used units with $\hbar=1$. We can see that if $a=2\alpha$ this is the same relation as that of equation (\ref{g2}) only upto a linear
term in $\alpha$. It can be shown that the smallest uncertainty in position occurs when $\langle \bf{p} \rangle$ $=0$ and $\Delta x_{min}=\frac{a}{2}$. The
momentum space wave function can be written as $\psi(p)=\langle p~|\psi \rangle $. On a dense domain in Hilbert space $\bf{x}$ and $\bf{p}$ act
as operators such that
\begin{equation}
\mathbf{p} ~\psi(p) = p ~\psi (p)~~,
\end{equation}
\begin{equation}
\label{ee14}
\mathbf{x} ~\psi (p) = i ~\Big[~(1-a ~p)~\frac{\partial}{\partial p}~\Big]~\psi(p)~~.
\end{equation}
This representation respects the commutation relation (\ref{ee12}) and the scalar product of two arbitrary wave functions in this representation is given by
\begin{equation}
\langle \phi ~|~ \psi \rangle = \int_{-\infty}^\infty ~dp~ \phi^*(p)~\psi (p)~~.
\end{equation}
Considering the standard derivation of the uncertainty relation we can see that if the state $|\psi \rangle $ obeys 
$\Delta x \Delta p = \frac{|\langle [x,p]\rangle|}{2}$ then it will obey the relation
\begin{equation}
\label{e16}
\Big(~\mathbf{x} - \langle \mathbf{x}\rangle + \frac{\langle [\mathbf{x},\mathbf{p}]\rangle}{2(\Delta p)^2}~(\mathbf{p}~ - \langle \mathbf{p}\rangle)\Big)~ 
|\psi \rangle = 0 ~~.
\end{equation}
The states of absolutely maximal localization can only be obtained for $\langle \mathbf{p}\rangle = 0$ with critical momentum uncertainty
$\Delta = \frac{2}{a}$. With equation (\ref{ee14}) and (\ref{e16}) we can calculate these states in momentum space and the states are
\begin{equation}
\psi_{\langle x \rangle} (p) = {\cal N} ~(1-a ~ p)^{i\frac{\langle x \rangle}{a}}~ e^{-\frac{p}{4}}~(1-a ~ p)^{-\frac{1}{4 a}}~~.
\end{equation}
Normalization of this wave function cannot be done as the integral required for this diverges. $(1-a ~ p)$ contains the first two terms of the
series form of $e^{-a p}$. As we have mentioned earlier that our approach is in some sense perturbative, so here we use an approximation
$(1-a ~ p) \approx e^{-a p}$. Using this we get
\begin{equation}
\psi_{\langle x \rangle} (p) = {\cal N}~ e^{i\langle x \rangle p} ~~.
\end{equation}
Now we can use a delta-function normalization and get 
\begin{equation}
\psi_{\langle x \rangle} (p) = \frac{1}{\sqrt{2\pi}}~ e^{i\langle x \rangle p} ~~,
\end{equation}
where we have used the relation
\begin{equation}
\langle \phi_{\langle x' \rangle}|~\psi_{\langle x \rangle}\rangle = \int_{-\infty}^{\infty}~\phi_{\langle x' \rangle}^*(p)~\psi_{\langle x \rangle}(p)~dp~
= \delta (\langle x' \rangle - \langle x \rangle) ~~.
\end{equation}
We could have also used the idea of box normalization for our purpose. The maximal localization states for a deformed Heisenberg algebra with a
linear term in $p$ in the commutation relation is a serious issue because the normalization is not possible. If we now apply equation (\ref{ee12}) to the Schr$\ddot{o}$dinger equation for a Harmonic Oscillator we simply get the MUP modified equation for the Oscillator as
\begin{equation}
\label{e22}
(1-a p)^2~\frac{\partial^2 \psi(p)}{\partial p^2} - a (1-a p)~\frac{\partial \psi(p)}{\partial p} + \frac{1}{m^2 \omega^2} (2 m E - p^2)~\psi (p) = 0 ~~.
\end{equation}
The solution is known in terms of associated Laguerre polynomials ${\cal L}_n^{2k}(\xi)$ \cite{bell,z17} and the modified energy eigen-values can be written as
\begin{equation}
\label{HOEV}
E_n = \left(n+\frac{1}{2}\right)~\omega - \frac{m ~\omega^2 ~a^2}{2} \left(n+\frac{1}{2}\right)^2 ~~.
\end{equation}
Here we have chosen $\hbar =1$. The spectrum is MUP corrected by the second term which will be treated as correction and we set an upper limit for $n$ as
\begin{equation}
n_{max} = \frac{1}{m~\omega ~a^2} ~~.
\end{equation}
Here we get a cut-off in the energy spectra which depends on our length scale and this can act as a regulator in the renormalization technique. This equation is exactly similar to equation (2.10) of \cite{poly} and (5.10) of \cite{asht} and hence the physical interpretation remain same except the length scale.
We can compare our results with those found in Polymer Quantization and put a bound on the commutator deformation parameter of MUP which
is $l_0$ where $a = \frac{2l_0~l_{pl}}{\hbar}$. The oscillations (vibrational) of a carbon monoxide molecule can be described by a harmonic oscillator with
mass $10^{-26}~kg$ and frequency $10^{15}~Hz$. With this and following the results of \cite{asht} we get an upper bound for $l_0$ which is given by
\begin{equation}
l_0 < 10^{16}~~.
\end{equation}
This is really very interesting as the measurement of the tunneling current in a scanning tunneling microscope for the simple harmonic oscillator also
sets the upper bound for $l_0$ to be $10^{17}$ \cite{bound}. Recently authors in \cite{nature} proposed an experimental scheme which is also within the reach of present technology to set bounds on the commutator deformation parameters like $l_0$. 
\par
We can now calculate the thermodynamic entities with the MUP corrected energy eigenvalue equation {\ref{HOEV}) for the Harmonic Oscillator. The partition function can be evaluated as
\begin{equation}
\label{hot1}
Z(\beta)=\sum_{n=0}^{n_{max}}\exp\left\{-\beta\left[\left(n+\frac{1}{2}\right)\omega-\frac{m\omega^{2}a^{2}}{2}\left(n+\frac{1}{2}\right)^{2}\right]\right\}~~.
\end{equation}
This sum can be evaluated but we do this in a perturbative sense to distinguish the first term as the partition function of a Schr\"odinger Harmonic Oscillator. We rewrite equation (\ref{hot1}) as
\begin{equation}
\label{hot2}
Z(\beta){\simeq} \frac{e^{-\frac{\beta \omega}{2}}}{(1- e^{- \beta \omega})} \left[ 1 + \frac{\beta m \omega^2 a^2}{8} \frac{(1 + e^{- 2 \beta \omega} + 
6 e^{-\beta \omega})}{(1 - e^{-\beta \omega})^2}\right] ~~.
\end{equation}
This equation guides us to write the MUP modified Helmholtz free energy as
\begin{equation}
\label{hot3}
F{\simeq}\frac{\cal N}{\beta}\left[\frac{\beta\omega}{2}+{\ln}\left(1-e^{-\beta\omega}\right)-\frac{a^{2}{\beta}m\omega^{2}}{8}\frac{\left(1+e^{-2\beta\omega}+6e^{-\beta\omega}\right)}{\left(1-e^{-\beta\omega}\right)^{2}}\right]~~.
\end{equation}
We now simply write the expressions for the entropy, internal energy and heat capacity respectively as
\begin{equation}
\label{hot4}
S={\cal N}{K_B}\left[\frac{\omega\beta}{\left(e^{\beta\omega}-1\right)}-\ln\left(1-e^{-\beta\omega}\right)+\frac{a^{2}m\omega^{3}\beta^{2}e^{\beta\omega}\left(e^{\beta\omega}+1\right)}{\left(e^{\beta\omega}-1\right)^{3}}\right]~~,
\end{equation}
\begin{equation}
\label{hot5}
U={\cal N}\omega\left[\frac{1}{2}+\frac{1}{\left(e^{\beta\omega}-1\right)}-\frac{a^{2}m\omega}{8\left(e^{\beta\omega}-1\right)^{3}}\left\{e^{3\beta\omega}+\left(5-8\beta\omega\right)e^{2\beta\omega}-\left(5+8\beta\omega\right)e^{\beta\omega}-1\right\}
\right]~~\&
\end{equation}
\begin{equation}
\label{hot6}
C_V=\frac{{\cal N}{K_B}\left(\omega\beta\right)^{2}}{\left(e^{\beta\omega}-1\right)^{2}}{\left(e^{\beta\omega}\right)}\left[1+\frac{a^{2}m\omega}{\left(e^{\beta\omega}-1\right)^{2}}\left\{2+\beta\omega\left(4e^{\beta \omega}+1\right)+e^{2\beta\omega}\left(\beta\omega-2\right)\right\}\right]~~.
\end{equation}
Here also we find that equations (\ref{hot4}), (\ref{hot5}) and (\ref{hot6}) resemble equations (3.13), (3.14) and (3.15) of \cite{poly}. We can clearly notice that in equations (\ref{e12}) and (\ref{hot4}) the MUP corrected entropy is increased. This may surprise us because the presence of a minimum length in the theory should reduce the number of microstates within a definite volume which would reduce the entropy. But here we would like to argue that when we calculate the the entropy we generally do not consider the effect of gravity or specifically neglect the gravitational degrees of freedom. Considering the effect of gravitational interaction in a theory will in turn induce a minimum length scale. In our case we have tried to incorporate the strong effect of gravity through the linear term in Planck length ($l_{pl}=\sqrt{G}$) in the modified uncertainty relation. Now increased degrees of freedom might contribute positively to the entropy. Though we believe that this is a naive argument but still we found additive corrections contributing positively in the case of black hole entropy also \cite{areahole}. 

\section*{Discussion}
In this short discussion we investigated two quantum systems (particle in a box and harmonic oscillator) in the MUP framework. We calculated the eigenvalues for
each and compared our results with the results from the polymer framework \cite{poly} and the GUP framework \cite{p58}. We have assumed that the quantum particles follow Maxwell-Boltzmann statistics with no spin. Interestingly we found that inspite of the presence of the linear term of the minimal length in the commutation relation with respect to the GUP, our eigenvalues match exactly but upto a prefactor sign of the first order correction. If we compare our results with the results found in polymer systems, they match exactly (approximate spectrum) although the length scales are of theoretically different origins. 
But the length scale which we get after comparing our result with those of Polymer, is almost similar as predicted in \cite{bound} for the simple harmonic oscillator. Later we also studied the statistical thermodynamics of both the systems and the MUP corrected thermodynamic entities are similar to those of polymer systems in terms of construction. Polymer Quantum Mechanics has originated from Loop Quantum Gravity and the later promises non-singular
cosmological models and a strong microscopic basis of black hole thermodynamics. So exploring relations, if any, in the basic foundations of the MUP and LQG
is worth interesting. An early indication about the possibility discussed above was discussed earlier by the author in \cite{kal}. 

\section*{Acknowledgements}
BM is very much thankful to Prof. Narayan Banerjee for guidance and support. SS acknowledges the fellowship given by Kishore Vaigyanik Protsahan Yojana with Reg.No. SB-1081267 (2008). The authors would also like to thank an anonymous referee for helpful comments and enlightening suggestions which immensely helped us improve the manuscript.

\end{document}